\documentclass{ita}
\usepackage[latin1]{inputenc}

\usepackage{amssymb}

\newcommand{\fa}{\forall}
\newcommand{\Ga}{\Gamma}
\newcommand{\Gas}{\Gamma^\star}
\newcommand{\Si}{\Sigma}
\newcommand{\Sis}{\Sigma^\star}
\newcommand{\Sio}{\Sigma^\omega}
\newcommand{\ra}{\rightarrow}
\newcommand{\hs}{\hspace{12mm}

\noi}
\newcommand{\lra}{\leftrightarrow}

\newcommand{\ol}{ $\omega$-language}
\newcommand{\orl}{ $\omega$-regular language}
\newcommand{\om}{\omega}
\newcommand{\nl}{\newline}
\newcommand{\noi}{\noindent}

\begin{document}

\title{On the Topological Complexity of \\ Infinitary Rational Relations }
\runningtitle{ On the Topological Complexity of Infinitary Rational
Relations }

\author{Olivier Finkel}
\address{Equipe de Logique Math\'ematique,
  U.F.R. de Math\'ematiques, Universit\'e Paris 7,  2 Place Jussieu, 75251
Paris
 cedex 05, France;
\email{finkel@logique.jussieu.fr}}
\date{}


\begin{abstract}
\noi We prove in this paper that there exists some infinitary rational
relations
which are analytic but non Borel sets, giving
 an answer to a question of Simonnet \cite{sim}.
\end{abstract}

\maketitle

\section{Introduction}

\noi Acceptance of infinite words by finite automata was firstly
considered by B\"uchi in order to study decidability of the monadic second
order theory
of one successor over the integers \cite{bu62}. Then the so called \orl s
have been
intensively studied and many applications have been found, see \cite{tho}
\cite{sta}
\cite{pp} for many results and references.
\nl Rational relations on finite words were studied in the sixties and
played
a fundamental role in the study of families of context free languages
\cite{ber}.
Their extension to rational  relations on infinite words was firstly
investigated by Gire and Nivat \cite{gire1} \cite{gn}. Infinitary rational
relations
are subsets of  $\Si_1^\om \times \Si_2^\om$,  where
$\Si_1$ and  $\Si_2$ are finite alphabets, which are recognized by
B\"uchi transducers or by $2$-tape finite B\"uchi automata with asynchronous
reading heads (there exists an extension to subsets of
$\Si_1^\om \times \Si_2^\om\times \ldots \times \Si_n^\om$
recognized by $n$-tape B\"uchi automata, with
$\Si_1, \ldots, \Si_n$ some finite alphabets, but we shall not need  to
consider it).
Since then they have been much studied, in particular in connection with
the rational functions they may define, see for example  \cite{cg} \cite{bc}
\cite{sim} \cite{sta} \cite{pri} for many results and references.

\hs The question of the complexity of such relations on infinite words
naturally arises. A way to investigate the complexity of infinitary
rational relations is to consider
their topological complexity and particularly  to locate them with regard to
the Borel and the projective hierarchies.
It is well known that every \ol~ accepted by a Turing machine with a
B\"uchi or Muller acceptance condition is an analytic set,  \cite{sta},
thus
every infinitary rational relation is an analytic set.
Simonnet asked  in \cite{sim} whether there exists some infinitary rational
relation
which is an analytic but non  Borel set.
 We give in this paper a positive answer to this question showing that there
exists
some non Borel (and even ${\bf \Si^1_1}$-complete) infinitary rational
relation.
The paper is organized as follows. In section 2 we introduce the notion of
transducers and
of infinitary rational relations. In section 3 we recall definitions of
Borel
and analytic sets, and we prove our main result in section 4.

\section{Infinitary rational relations}

\noi Let $\Si$ be a finite alphabet whose elements are called letters.
A non-empty finite word over $\Si$ is a finite sequence of letters:
 $x=a_1a_2\ldots a_n$ where $\fa i\in [1; n]$ $a_i \in\Si$.
 We shall denote $x(i)=a_i$ the $i^{th}$ letter of $x$
and $x[i]=x(1)\ldots x(i)$ for $i\leq n$. The length of $x$ is $|x|=n$.
The empty word will be denoted by $\lambda$ and has 0 letter. Its length is
0.
 The set of finite words over $\Si$ is denoted $\Sis$.
 $\Si^+ = \Sis - \{\lambda\}$ is the set of non empty words over $\Si$.
 A (finitary) language $L$ over $\Si$ is a subset of $\Sis$.
 The usual concatenation product of $u$ and $v$ will be denoted by $u.v$ or
just  $uv$.
 For $V\subseteq \Sis$, we denote  \quad
$V^\star = \{ v_1\ldots v_n  /  n\in \mathbb{N}\quad and \quad v_i \in V
\quad \fa i \in [1; n] \}$.

\hs   The first infinite ordinal is $\om$.
An $\om$-word over $\Si$ is an $\om$ -sequence $a_1a_2\ldots a_n \ldots$,
where
$a_i \in\Sigma , \fa i\geq 1$.
 When $\sigma$ is an $\om$-word over $\Si$, we write
 $\sigma =\sigma(1)\sigma(2)\ldots  \sigma(n) \ldots $
and $\sigma[n]=\sigma(1)\sigma(2)\ldots  \sigma(n)$ the finite word of
length $n$,
prefix of $\sigma$.
The set of $\om$-words over  the alphabet $\Si$ is denoted by $\Si^\om$.
 An  $\om$-language over an alphabet $\Sigma$ is a subset of  $\Si^\om$.
For $V\subseteq \Sis$,
 $V^\om = \{ \sigma =u_1\ldots  u_n\ldots  \in \Si^\om /  u_i\in V, \fa
i\geq 1 \}$
is the $\om$-power of $V$.
 The concatenation product is extended to the product of a
finite word $u$ and an $\om$-word $v$:
the infinite word $u.v$ is then the $\om$-word such that:
 $(u.v)(k)=u(k)$  if $k\leq |u|$ , and  $(u.v)(k)=v(k-|u|)$  if $k>|u|$.
\nl The prefix relation is denoted $\sqsubseteq$: the finite word $u$ is a
prefix of the finite
word $v$ (respectively,  the infinite word $v$), denoted $u\sqsubseteq v$,
 if and only if there exists a finite word $w$
(respectively,  an infinite word $w$), such that $v=u.w$.

\hs  We assume the reader to be familiar with the theory of formal languages
and of
\orl s, see \cite{bu62} \cite{tho} \cite{eh} \cite{sta} \cite{pp}
for many results and references. We recall that \orl s form the class of \ol
s accepted
by finite automata with a   B\"uchi acceptance condition and this class is
the omega Kleene
closure of the class of regular finitary languages.

\hs  We are going now to introduce  the notion of infinitary rational
relation which
extends the notion of \orl , via definition by  B\"uchi transducers:

\begin{dfntn}
A B\"uchi transducer is a sextuple $\mathcal{T}=(K, \Si, \Ga, \Delta, q_0,
F)$, where
$K$ is a finite set of states, $\Si$ and $\Ga$ are finite sets called the
input and
the output alphabets,
$\Delta$ is a finite subset of $K \times \Sis \times \Gas \times K$ called
the set of transitions, $q_0$ is the initial state,  and $F \subseteq K$ is
the set of
accepting states.
\nl A computation $\mathcal{C}$ of the transducer $\mathcal{T}$ is an
infinite sequence of transitions
$$(q_0, u_1, v_1, q_1), (q_1, u_2, v_2, q_2), \ldots (q_{i-1}, u_{i}, v_{i},
q_{i}),
(q_i, u_{i+1}, v_{i+1}, q_{i+1}), \ldots $$
\noi The computation is said to be successful iff there exists a final state
$q_f \in F$
and infinitely many integers $i\geq 0$ such that $q_i=q_f$.
\nl The input word of the computation is $u=u_1.u_2.u_3 \ldots$
\nl The output word of the computation is $v=v_1.v_2.v_3 \ldots$
\nl Then the input and the output words may be finite or infinite.
\nl The infinitary rational relation $R(\mathcal{T})\subseteq \Sio \times
\Ga^\om$
recognized by the B\"uchi transducer $\mathcal{T}$
is the set of couples $(u, v) \in \Sio \times \Ga^\om$ such that $u$ and $v$
are the input
and the output words of some successful computation $\mathcal{C}$ of
$\mathcal{T}$.
\nl The set of infinitary rational relations will be denoted $RAT$.
\end{dfntn}

\begin{rmrk}\label{ol s}
An infinitary rational relation is a subset of $\Sio \times \Ga^\om$ for two
finite
alphabets $\Si$ and $\Ga$. One can also consider that it is an \ol~ over the
finite
alphabet $\Si \times \Ga$. If $(u, v) \in \Sio \times \Ga^\om$,
 one can consider this couple of
infinite words as a single infinite word $(u(1),v(1)).(u(2), v(2)).(u(3),
v(3))\ldots $
over the alphabet $\Si \times \Ga$.
 We shall use this fact to investigate the topological complexity
of infinitary rational relations.
 \end{rmrk}

\section{Borel and analytic sets}

\noi We assume the reader to be familiar with basic notions of topology
which
may be found in  \cite{ku} \cite{mos} \cite{kec} \cite{lt} \cite{sta}
\cite{pp}.
\nl For a finite alphabet $X$ having at least two letters we shall consider
$X^\om$
as a topological space with the Cantor topology.
The open sets of $X^\om$ are the sets in the form $W.X^\om$, where
$W\subseteq X^\star$.
A set $L\subseteq X^\om$ is a closed set iff its complement $X^\om - L$ is
an open set.
We define now the next classes of the Borel Hierarchy:

\begin{dfntn}
The classes ${\bf \Si_n^0}$ and ${\bf \Pi_n^0 }$ of the Borel Hierarchy
 on the topological space $X^\om$  are defined as follows:
\nl ${\bf \Si^0_1 }$ is the class of open sets of $X^\om$.
\nl ${\bf \Pi^0_1 }$ is the class of closed sets of $X^\om$.
\nl And for any integer $n\geq 1$:
\nl ${\bf \Si^0_{n+1} }$   is the class of countable unions
of ${\bf \Pi^0_n }$-subsets of  $X^\om$.
\nl ${\bf \Pi^0_{n+1} }$ is the class of countable intersections of
${\bf \Si^0_n}$-subsets of $X^\om$.
\nl The Borel Hierarchy is also defined for transfinite levels.
The classes ${\bf \Si^0_\alpha }$
 and ${\bf \Pi^0_\alpha }$, for a countable ordinal $\alpha$, are defined in
the
 following way:
\nl ${\bf \Si^0_\alpha }$ is the class of countable unions of subsets of
$X^\om$ in
$\cup_{\gamma <\alpha}{\bf \Pi^0_\gamma }$.
 \nl ${\bf \Pi^0_\alpha }$ is the class of countable intersections of
subsets of $X^\om$ in
$\cup_{\gamma <\alpha}{\bf \Si^0_\gamma }$.
\end{dfntn}

\noi
There are also some subsets of $X^\om$ which are not Borel sets.
In particular the class of Borel subsets of $X^\om$ is strictly included
into the class
${\bf \Si^1_1}$ of {\bf analytic} subsets of $X^\om$.
A subset $A$ of  $X^\om$ is an  {\bf analytic} set
iff there exists another finite set $Y$ and a Borel subset $B$  of
$(X\times Y)^\om$
such that $ x \in A \lra \exists y \in Y^\om $ such that $(x, y) \in B$,
where $(x, y)$ is the infinite word over the alphabet $X\times Y$ such that
$(x, y)(i)=(x(i),y(i))$ for each  integer $i\geq 1$.

\hs Recall also the notion of completeness with regard to reduction by
continuous functions.
If $\alpha$ be a countable ordinal, a set $F\subseteq X^\om$ is said to be
a ${\bf \Si^0_\alpha}$  (respectively ${\bf \Pi^0_\alpha}$, ${\bf
\Si^1_1}$)-complete set
iff for any set $E\subseteq Y^\om$  (with $Y$ a finite alphabet):
 $E\in {\bf \Si^0_\alpha}$ (respectively $E\in {\bf \Pi^0_\alpha}$, ${\bf
\Si^1_1}$)
iff there exists a continuous
function $f: Y^\om \ra X^\om$ such that $E = f^{-1}(F)$.
\nl  A ${\bf \Si^0_\alpha}$
 (respectively ${\bf \Pi^0_\alpha}$, ${\bf \Si^1_1}$)-complete set is a
${\bf \Si^0_\alpha}$
 (respectively ${\bf \Pi^0_\alpha}$, ${\bf \Si^1_1}$)- set which is
in some sense a set of the highest
topological complexity among the ${\bf \Si^0_\alpha}$
 (respectively ${\bf \Pi^0_\alpha}$, ${\bf \Si^1_1}$)- sets.
  ${\bf \Si^0_n}$
 (respectively ${\bf \Pi^0_n}$)-complete sets, with $n$ an integer $\geq 1$,
 are thoroughly characterized in \cite{stac}.

\hs  The $\om$-language $\mathcal{A}=(0^\star.1)^\om$
is a well known example of
${\bf \Pi^0_2 }$-complete set which will be used below. It is the set of
$\om$-words over the alphabet $\{0, 1\}$ with infinitely many occurrences of
the letter $1$.

\section{${\bf \Si^1_1}$-complete infinitary  rational relations}

\noi We can now state our main result:

\begin{thrm} There exists some ${\bf \Si^1_1}$-complete (hence non Borel)
infinitary  rational relations.
\end{thrm}

\begin{proof}
 We shall use here results about languages of infinite binary trees whose
nodes
are labelled in a finite alphabet $\Si$.
\nl A node of an infinite binary tree is represented by a finite  word over
the alphabet $\{l, r\}$ where $r$ means "right" and $l$ means "left". Then
an
infinite binary tree whose nodes are labelled  in $\Si$ is identified with a
function
$t: \{l, r\}^\star \ra \Si$. The set of  infinite binary trees labelled in
$\Si$ will be
denoted $T_\Si^\om$.

\hs  There is a natural topology on this set $T_\Si^\om$ \cite{mos},
\cite{lt}, \cite{sim}.
It is defined
by the following distance. Let $t$ and $s$ be two distinct infinite trees in
$T_\Si^\om$.
Then the distance between $t$ and $s$ is $\frac{1}{2^n}$ where $n$ is the
smallest integer
such that $t(x)\neq s(x)$ for some word $x\in \{l, r\}^\star$ of length $n$.
\nl The open sets are then in the form $T_0.T_\Si^\om$ where $T_0$ is a set
of finite labelled
trees. $T_0.T_\Si^\om$ is the set of infinite binary trees
which extend some finite labelled binary tree $t_0\in T_0$, $t_0$ is here a
sort of prefix,
an "initial subtree"
of a tree in $t_0.T_\Si^\om$.
\nl  For an alphabet $\Si$ having at least two letters the topological
space $T_\Si^\om$ is homeomorphic to the Cantor set $\Sio$.
Borel and analytic subsets of  $T_\Si^\om$ are defined from open
sets in the same manner as in the case of the topological space $\Si^\om$.

\hs Let $t$ be a tree. A branch $B$ of $t$ is a subset of the set of nodes
of $t$ which
is linearly ordered by the tree partial order $\sqsubseteq$
 and which is closed under prefix relation,
i.e. if  $x$ and $y$ are nodes of $t$ such that $y\in B$ and $x \sqsubseteq
y$ then $x\in B$.
\nl A branch $B$ of a tree is said to be maximal iff there is not any other
branch of $t$
which strictly contains $B$.

\hs Let $t$ be an infinite binary tree in $T_\Si^\om$. If $B$ is a maximal
branch of $t$,
then this branch is infinite. Let $(u_i)_{i\geq 0}$ be the enumeration of
the nodes in $B$
which is strictly increasing for the prefix order.
\nl  The infinite sequence of labels of the nodes of  such a maximal
branch $B$, i.e. $t(u_0)t(u_1)....t(u_n).....$  is called a path. It is an
$\om$-word
over the alphabet $\Si$.

\hs Let then $L\subseteq \Si^\om$ be an \ol~ over $\Si$. We denote $Path(L)$
the set of
infinite trees $t$ in $T_\Si^\om$ such that $t$ has at least one path in
$L$.

\hs It is well known that if $L\subseteq \Si^\om$ is an \ol~ over $\Si$
which is a
${\bf \Pi^0_2 }$-complete subset of $\Si^\om$ (or a Borel
set of higher complexity in the Borel
hierarchy) then the set $Path(L)$  is a ${\bf \Si^1_1 }$-complete subset of
$T_\Si^\om$.
Hence in particular $Path(L)$  is not a Borel set, \cite{niw85}
\cite{simcras} \cite{sim}.

\hs Whenever $B \subseteq \Sio$ is a regular \ol , we shall find a
rational relation $R \subseteq (\Si \cup\{A\})^\om \times (\Si
\cup\{A\})^\om$
 and a continuous function

$$h: T_\Si^\om   \ra  ((\Si\cup\{A\})\times (\Si \cup\{A\}))^\om $$
\noi such that  $Path(B) = h^{-1} ( R )$.
For that we shall code trees labelled in $\Si$ by  words over the finite
alphabet
$(\Si\cup\{A\})\times (\Si \cup\{A\})$ where
$A$ is  supposed to be a new letter not in $\Si$.

\hs Consider now the set $\{l, r\}^\star$ of nodes of binary infinite trees.
For each integer $n\geq 0$, call $C_n$ the set of words of length $n$ of
$\{l, r\}^\star$.
Then $C_0=\{\lambda\}$, $C_1=\{l, r\}$, $C_2=\{ll, lr, rl, rr\}$ and so on.
$C_n$ is the set of nodes which appear in the $(n+1)^{th}$ level of an
infinite binary tree.
The number of nodes of $C_n$ is $card(C_n)=2^n$.  We consider now
the lexicographic order on $C_n$ (assuming that $l$ is before $r$ for this
order).
Then, in the enumeration of the nodes with regard to this order, the  nodes
of $C_1$ will
be: $l, r$; the nodes of $C_3$ will be: $ lll, llr, lrl, lrr, rll, rlr, rrl,
rrr$.
\nl Let $u^n_1,..., u^n_j,..., u^n_{2^n}$ be such an enumeration of $C_n$ in
the lexicographic order and let $v^n_1,..., v^n_j,..., v^n_{2^n}$ be the
enumeration of the
elements
of $C_n$ in the reverse order. Then for all integers $n\geq 0$ and $i$,
$1\leq i\leq 2^n$,
it holds that $v_i^n=u^n_{2^n+1-i}$.

\hs We define now the code of a tree $t$ in $ T_\Si^\om$.
Let $A$ be a new letter not in $\Si$. The code of the tree $t$ is
an $\om$-word $\sigma$ over the alphabet  $(\Si\cup\{A\}) \times
(\Si\cup\{A\})$
which may be written in the form $(\sigma_1, \sigma_2)$, where $\sigma_1$
and $\sigma_2$ are
$\om$-words over the alphabet $(\Si\cup\{A\})$.

\hs The $\om$-word $\sigma_1$  enumerates the  labels of the nodes
of the tree $t$ which appear at levels $1$, $3$, $5$, \ldots, $2n+1$,
\ldots,
i.e. at odd levels. More precisely the word $\sigma_1$ begins with the label
$t(v_1^0)$
of the node at level $1$, followed by an $A$, followed by the labels of the
nodes
of the third level enumerated in the {\bf reverse lexicographic order}, i.e.
$t(v_1^2)t(v_2^2)t(v_3^2)t(v_4^2)$, followed by an $A$, followed by
the labels of the nodes
of the $5^{th}$  level enumerated in the reverse lexicographic order, i.e.
 $t(v_1^4)t(v_2^4)t(v_3^4)\ldots t(v_{16}^4)$, and so on \ldots
\nl For each integer $n\geq 0$, the labels of the nodes of $C_{2n}$,
enumerated in the
{\bf reverse lexicographic order}, are placed  before
those of $C_{2n+2}$ and these two sets of labels are separated by an $A$.

\hs The construction of the $\om$-word $\sigma_2$ is very similar
but it successively enumerates,
in the {\bf lexicographic order}, the labels of nodes occuring at even
levels. So the word $\sigma_2$
is in the form
$$\sigma_2 = t(u_1^1)t(u_2^1)A
t(u_1^3)t(u_2^3)t(u_3^3)t(u_4^3)t(u_5^3)t(u_6^3)t(u_7^3)t(u_8^3)A \ldots$$

\noi For each integer $n\geq 0$, the labels of the nodes of $C_{2n+1}$ are
enumerated before
those of $C_{2n+3}$ and these two
sets of labels are separated by an $A$. Moreover the labels of the nodes of
$C_{2n+1}$, for
$n\geq 0$, are enumerated  in the {\bf lexicographic order} (for the nodes).

 \hs Let then $h$ be the mapping from $T_\Si^\om$ into
 $((\Si\cup\{A\})\times (\Si \cup\{A\}))^\om$
such that for every labelled binary infinite tree $t$ of $T_\Si^\om$,
$h(t)$ is the code $(\sigma_1, \sigma_2)$ of the tree as defined above.
It is easy to see, from the definition of $h$ and of the  order of the
enumeration
of labels of nodes (they are enumerated level after level in the increasing
order),
that $h$ is a continuous function from $T_\Si^\om$ into
$((\Si\cup\{A\})\times (\Si \cup\{A\}))^\om$.

\hs Now we are looking for a rational relation $R$  such that
for every  tree $t \in T_\Si^\om$,  $h(t) \in R$ if and only if  $t$ has
a path in  $B$. Then  we shall have $Path(B) = h^{-1} ( R )$.

\hs We shall first describe the rational relation $R$ which  is an \ol~ over
the alphabet
$((\Si\cup\{A\})\times (\Si \cup\{A\}))$.  Every word of $R$ may be seen as
a couple
$y=(y_1, y_2)$ of $\om$-words over the alphabet $\Si\cup\{A\}$. Now $y=(y_1,
y_2)$ is in $R$
if and only if it is in the form

\hs $y_1 = x(1).u_1.A.v_2.x(3).u_3.A.v_4.x(5).u_5.A. \ldots
A.v_{2n}.x(2n+1).u_{2n+1}.A\ldots$
\nl $y_2 = ~~~ v_1.x(2).u_2.A.v_3.x(4).u_4.A. \ldots
A.v_{2n+1}.x(2n+2).u_{2n+2}.A\ldots$

\hs where for all integers $i\geq 1$, $x(i)\in \Si$ and  $u_i, v_i \in \Sis$
and
$$|v_i|=2|u_i|  ~~~~ \mbox{  or }  ~~~~|v_i|=2|u_i|+1$$
and the $\om$-word $x=x(1)x(2)\ldots x(n)\ldots $ is in $B$.

\hs If such an $\om$-word $y=(y_1, y_2)$ is the code $h(t)$ of a tree $t \in
T_\Si^\om$,
then $x(1)=t(v_1^0)$ and $u_1=\lambda$, then $|v_1|=2|u_1|=0$ or
$|v_1|=2|u_1|+1=1$. Therefore
if  $|v_1|=0$ then $x(2)=t(u_1^1)$ and if $|v_1|=1$ then $x(2)=t(u_2^1)$.
Then the choice of
 $|v_1|=2|u_1|$ or of $|v_1|=2|u_1|+1$ implies that $x(2)$ is the label
of the left or the rigth successor of the root node $v_1^0=\lambda$.

\hs By construction this phenomenon will happen for further levels. The
choice of
$|v_i|=2|u_i|$ or of $|v_i|=2|u_i|+1$ determines one of the two successor
nodes
of a node at level $i$ thus the successive choices determine a branch of $t$
and the labels of nodes of this branch form a path $x(1)x(2)x(3)\ldots x(n)
\ldots $
which is in $B$. Thus for a tree $t \in T_\Si^\om$,  $h(t)\in R$ if and only
if
$t \in Path(B)$ then $Path(B) = h^{-1} ( R )$.
\nl Remark that $R$ does not contain only codes of trees but such a code
$h(t)$ is in $R$
iff  $t\in Path(B)$ and this fact  suffices for our proof.

\hs  Hence if  $B$ is a Borel set which is  a ${\bf \Pi_2^0}$-complete
subset
of $\Si^\om$ (or a set of higher complexity in the Borel hierarchy), the
language
$h^{-1}(R)=Path(B)$ is a ${\bf \Si^1_1}$-complete subset of $T_\Si^\om$.
Then the \ol~  $R$ is at least  ${\bf \Si^1_1}$-complete because $h$
is a continuous function.
 \nl Note that here $h$ is a continuous function:
$T_\Si^\om \ra ((\Si\cup\{A\})\times (\Si \cup\{A\}))^\om$
and the preceding definition of ${\bf \Si^1_1}$-complete set involves
continuous reductions:
 $X^\om \ra Y^\om$; but the two topological spaces $T_\Si^\om $ and $Y^\om$
have
good similar properties (they are zero-dimensional polish spaces,
see \cite{pp} \cite{kec} \cite{sim}, in fact they are homeomorphic)
which enable to extend the previous definition  to this new case.
\nl Indeed  $R$  is  a ${\bf \Si^1_1}$-complete subset of
$((\Si\cup\{A\})\times (\Si \cup\{A\}))^\om$
because every infinitary rational relation is a ${\bf \Si^1_1}$-set.
\nl Then in that case $R$ is not a Borel set because a ${\bf
\Si^1_1}$-complete
set is not a Borel set. This gives infinitely many non Borel infinitary
rational relations,
because there exist infinitely many  ${\bf \Pi_2^0}$-complete \orl s.

\hs It remains to show that if $B$ is an \orl~ then $R$ is an infinitary
rational relation.
In fact this is easy to see from the definition of $R$.  We shall
explicitely give
a B\"uchi transducer defining $R$ in the following simple case: $\Si =\{0,
1\}$ and
$B=(0^\star.1)^\om$ is a well known example of ${\bf \Pi_2^0}$-complete \orl
.

\hs The infinitary rational relation $R$ is then recognized by the following
B\"uchi transducer  $\mathcal{T}=(K, (\Si \cup \{A\}), (\Si \cup \{A\}),
\Delta, q_0, F)$,
where
$$K=\{q_0, q_1, q_2, q_3, q_4, q_1^0, q_1^1, q_2^0, q_2^1\}$$
\noi  is a finite set of states,
$\{0, 1, A\}$ is the input {\it and} the output alphabet,
$q_0$ is the initial state,  and $F = \{q_1^1, q_2^1\}$ is the set of
accepting states.
Moreover $\Delta \subseteq K \times
(\Si \cup \{A\})^\star \times (\Si \cup \{A\})^\star \times K$ is the finite
 set of transitions, containing  the following transitions:

\hs $(q_0, 0, \lambda, q_1)$ and $(q_0, 1, \lambda, q_1)$,
\nl  $(q_1, u, v, q_1)$, for all words $u, v \in \Sis$ with $|u|=1$ and
$|v|=2$,
\nl  $(q_1, \lambda, v, q_2)$,  for $v\in \{0, 1, \lambda\}$,
\nl  $(q_2, A, 0, q_1^0)$ and  $(q_2, A, 1, q_1^1)$,
\nl  $(q, u, v, q_3)$,  for all $u, v \in \Sis$ with $|u|=2$ and $|v|=1$ and
$q\in \{q_1^0, q_1^1, q_3\}$,
\nl  $(q, u, \lambda, q_4)$,  for  $u \in \{0, 1, \lambda\}$  and
$q\in \{q_1^0, q_1^1, q_3\}$,
\nl  $(q_4, 0, A, q_2^0)$ and   $(q_4, 1, A, q_2^1)$,
\nl  $(q_2^0, \lambda, \lambda, q_1)$ and   $(q_2^1, \lambda, \lambda,
q_1)$.
\end{proof}

\begin{rmrk} We could of course have avoided the set of transitions to
contain
some transitions with both the input and the output words being  empty, like
the two last ones:
$(q_2^0, \lambda, \lambda, q_1)$ and   $(q_2^1, \lambda, \lambda, q_1)$.
\end{rmrk}

\begin{rmrk} We have shown that there exists some infinitary  rational
relations
 which are  ${\bf \Si^1_1}$-complete hence non Borel. In particular this
implies that these
infinitary  rational relations are not arithmetical sets because every
arithmetical set
is a Borel set (of finite rank). We refer to \cite{stac} \cite{sta}
for definitions and results about the
arithmetical hierarchy over  sets of infinite words over a finite alphabet
$\Si$.
\end{rmrk}

\begin{rmrk} From the preceding example we can easily find a ${\bf
\Si^1_1}$-complete
infinitary  rational relation in the form $S^\om$ where $S$ is a rational
relation
over finite words, see \cite{ber} \cite{gire1} \cite{pri} about finitary
rational relations.
\end{rmrk}

\noi {\bf  Acknowledgements.} Thanks to Jean-Pierre Ressayre and  Pierre
Simonnet
for useful discussions.

\begin{footnotesize}

\end{footnotesize}

\end{document}